\documentclass[twocolumn,amsmath,amssymb,aps,pra,10pt,showkeys,superscriptaddress,longbibliography
,nofootinbib
]{revtex4-2}
\usepackage{dcolumn}
\usepackage{bm}
\usepackage{mathtools, cases}
\usepackage[unicode]{hyperref}
\usepackage{graphicx}
\usepackage{times,xcolor}
\usepackage{soul}
\hypersetup{colorlinks=true,linkcolor=blue,citecolor=blue,filecolor=black,urlcolor=blue}

\begin{document}

\title{Dense Random Packing of Disks With a Power-Law Size Distribution in Thermodynamic Limit}

\author{Alexander Yu. Cherny}
\affiliation{Joint Institute for Nuclear Research, Dubna 141980, Russian Federation}

\author{Eugen M. Anitas}
\email[Corresponding author, e-mail:~]{anitas@theor.jinr.ru}
\affiliation{Joint Institute for Nuclear Research, Dubna 141980, Russian Federation}
\affiliation{Horia Hulubei, National Institute of Physics and Nuclear Engineering, RO-077125 Bucharest-Magurele, Romania}

\author{Artem A. Vladimirov}
\affiliation{Joint Institute for Nuclear Research, Dubna 141980, Russian Federation}

\author{Vladimir A. Osipov}
\affiliation{Joint Institute for Nuclear Research, Dubna 141980, Russian Federation}

\date{\today}

\begin{abstract}
The correlation properties of a random system of densely packed disks, obeying a power-law size distribution, are analyzed in reciprocal space in the thermodynamic limit. This limit assumes that the total number of disks increases infinitely, while the mean density of the disk centers and the range of the size distribution are kept constant. We investigate the structure factor dependence on momentum transfer across various number of disks and extrapolate these findings to the thermodynamic limit. The fractal power-law decay of the structure factor is recovered in reciprocal space within the fractal range, which corresponds to the range of the size distribution in real space. The fractal exponent coincides with the exponent of the power-law size distribution as was shown previously by the authors [\href{https://doi.org/10.1063/5.0134813}{A. Yu. Cherny, E. M. Anitas, V. A. Osipov, \textit{J. Chem. Phys.} \textbf{158}(4), 044114 (2023)}]. The dependence of the structure factor on density is examined. As is found, the power-law exponent remains unchanged but the fractal range shrinks when the packing fraction decreases. Additionally, the finite-size effects are studied at extremely low momenta of the order of the inverse system size. We show that the structure factor is parabolic in this region and calculate the prefactor analytically. The obtained results reveal fractal-like properties of the packing and can be used to analyze small-angle scattering from such systems.
\end{abstract}


\maketitle

\section{Introduction}
\label{sec:intro}
Dense random packings of granular particles or pores with a power-law size distribution are of significant importance in various applications  \cite{torquato18}, including emulsion \cite{kwok20}, soil \cite{iglauer10}, pavements \cite{patino20}, ceramics \cite{bhatta21}, and concrete technologies \cite{qifeng21}. Specifically, in hard materials, power-law distributions typically yield the highest densities, which are crucial for obtaining  exceptional resistance \cite{hermann03}. Extensive numerical and analytical studies were conducted on these systems. These studies range from modeling fragmentation of grains in 2D packings \cite{astrom98} to examining 3D samples composed of spherical grains~\cite{patino20} as well as general nonoverlapping circle, sphere, and drop packings \cite{aste96}. Thus, understanding the structural and correlation properties of dense packings has direct implications for improving the design and efficiency of materials used in construction, manufacturing, and environmental applications.

Besides practical applications, there is a fundamental challenge in understanding the correlation properties of dense random packings of particles with a power-law size distribution. These packings exhibit complex, fractal-like properties \cite{cherny22,monti22,cherny23}, which are controlled by two parameters: the range of the power-law size distribution and its exponent. The correlation properties are quantitatively described in terms of these parameters: the higher the range of power-law distribution, the lengthier the fractal ranges of mass-radius relation, pair correlation function, and structure factor. In these regions, we observed \cite{cherny23} that the power-law and fractal exponents coincide.

The dense packing limit used in the previous paper \cite{cherny23} assumes that the total number of the disks tends to infinity but the area occupied by the system remains finite. In this limit, the smallest size of these disks tends to zero, which implies that the largest-to-smallest size ratio increases infinitely. In this paper, we extend our results \cite{cherny23} to a more realistic scenario involving a macroscopic number of disks and a fixed size ratio. Thus, we consider the structure factor as a function of momentum transfer in the thermodynamic limit when the area $A$ and the number of disks $N$ increase while the density $A/N$ and the size ratio remain constant.

In statistical physics, it is well-established that the boundary conditions do not play a role in the thermodynamic limit when calculating the main asymptotics of physical quantities (see Sec.~\ref{sec:alg}), and we can use any of them we might need.
In this paper, the positions of disks are restricted by external impenetrable borders, which is equivalent to the zero boundary conditions (see Sec.~\ref{sec:alg} below).

To this end, we introduce a time-efficient algorithm for generating dense random packing of disks with a given size distribution using Delaunay triangulation (DT). \cite{del34}. The triangulation is widely employed in compact packing problems for disks in 2D and spheres in 3D~\cite{reboul08} as well as in analyzing pore spaces in porous media \cite{nguyen21}. For the dense packing, the most time-consuming task is to check for possible intersections between a newly placed disk and a set of $N$ already placed disks. The DT significantly optimizes the computation time, requiring only $O (\log N)$ repeated operations to select neighbours of the newly placed disk and calculate distances to them. By contrast, the packing requires $O(N)$ operations without the
DT. Thus, the total number of operations for dense packing with and without DT is $O(N\log N)$ and $O(N^{2})$, respectively. For validation of the DT-based algorithm, we also generate the packings with the random sequential addition (RSA) algorithm and compare the results. A modified version of the RSA algorithm is considered and discussed in Sec.~\ref{sec:alg} below.

In addition, we study finite-size effects for the structure factor $S(q)$ at extremely low momentum transfers $q\lesssim 2\pi/s$, where $s$ is the edge size of the embedding square. If the structure factor is defined through the density \emph{fluctuations}, then $S(q)$ is proportional to $q^{2}$ within this range. As is shown in Sec.~\ref{smallQ} below, the prefactor is determined by the variance of the disk positions, which are randomly distributed over the area.

\section{Algorithms of disk fractal generation}
\label{sec:alg}

We consider a set of $N$ disks obeying a power-law distribution with the exponent $D$~\cite{cherny22,cherny23}. In two dimensions, the exponent satisfies the condition $0<D<2$. The number of disks $d N(r)$ whose radii fall within the range ($r,r + d r$) is proportional to $dr/r^{D+1}$. The radii vary within the range of the disrtibution, from $a$ to $R$. The size ratio $R/a$ determines the length of the range of the distribution on a log-scale.

\subsection{Random-sequential-addition grid}
\label{RSA}

The RSA algorithm is described in detail in Ref.~\cite{cherny23}. For a large number of particles, the algorithm becomes excessively time-consuming, requiring approximately $O(N^2)$ operations, see Appendix \ref{sec:RSAcompl}.  To reduce the computation time, we adopt a simplified approach and build the entire system from $k^2$ randomly generated cells. The system is a square grid of $k$ by $k$ cells, thereafter called RSA grid. Each cell, created with the RSA method \cite{cherny23}, consists of $M=N/k^2$ disks of radii $r_i = R\, {i}^{-1/D}$,  $i = 1, ..., M$. Hence, the total number of disks in the grid is equal to $N=k^2 (R/a)^{D}$.

The thermodynamic limit is simulated by $k\to\infty$ with $M$ and $R/a$ being constant. This scheme enables us to reduce the number of operations to $O(M^2k^2)=O(N/k^2)$. It artificially  changes the spatial long-range correlations of the disk positions at the distances bigger than $s/k$, but the fractal range remains intact (see  Sec.~\ref{sec:FR} below).

Figure \ref{fig1} (Top) shows a grid of $5 \times 5$ cells with each cell containing $M = 5000$ disks with $D = 1.5$ and $R/a\simeq 3.4\times 10^{-3}$. For these control parameters, the packing fraction is about $95\%$ and $s/R \simeq 15$. Disks of the same radius are uniformly colored; the largest disks are red, while the smallest are purple.

\begin{figure}[tb]
\includegraphics[width=0.6\columnwidth]{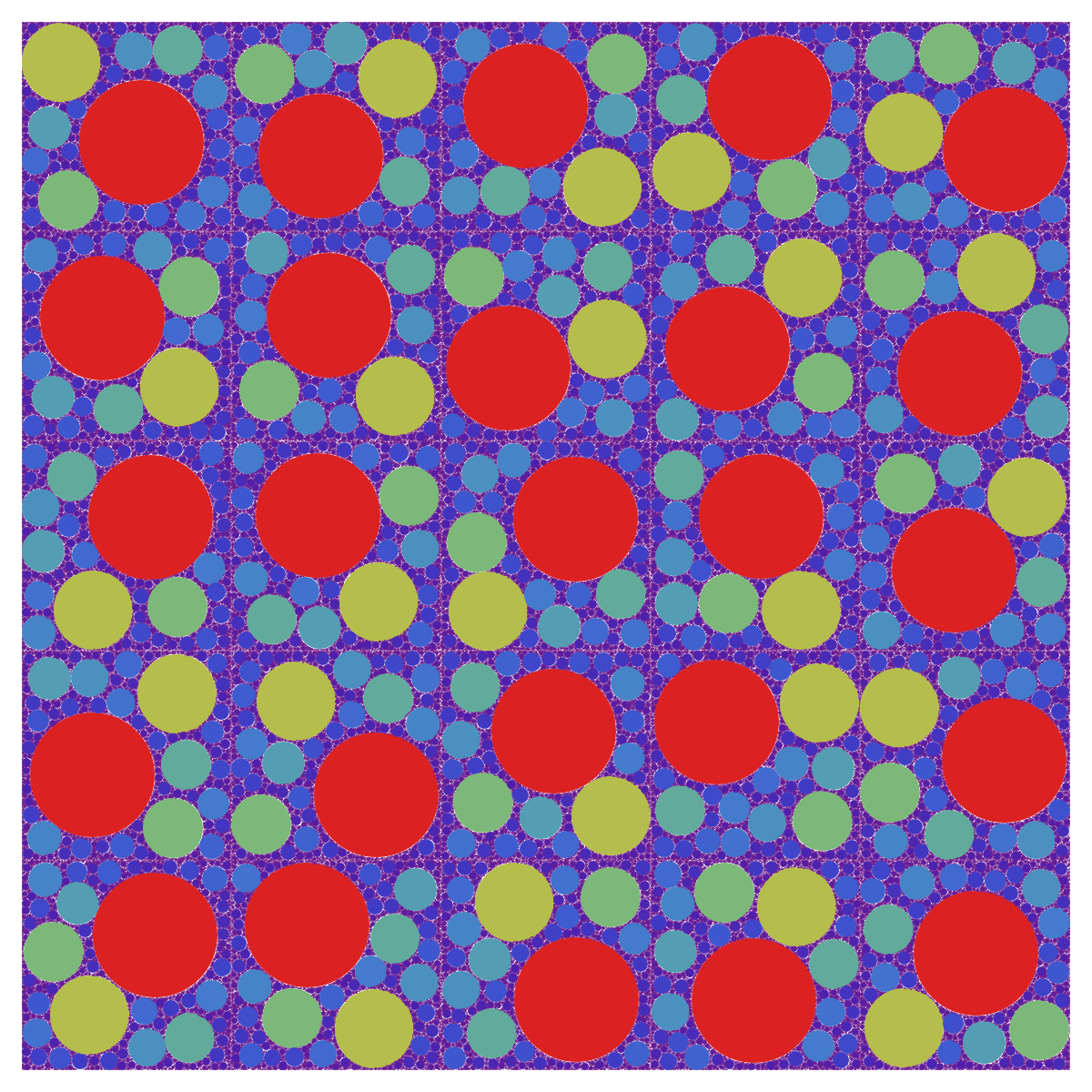}
\includegraphics[width=0.6\columnwidth]{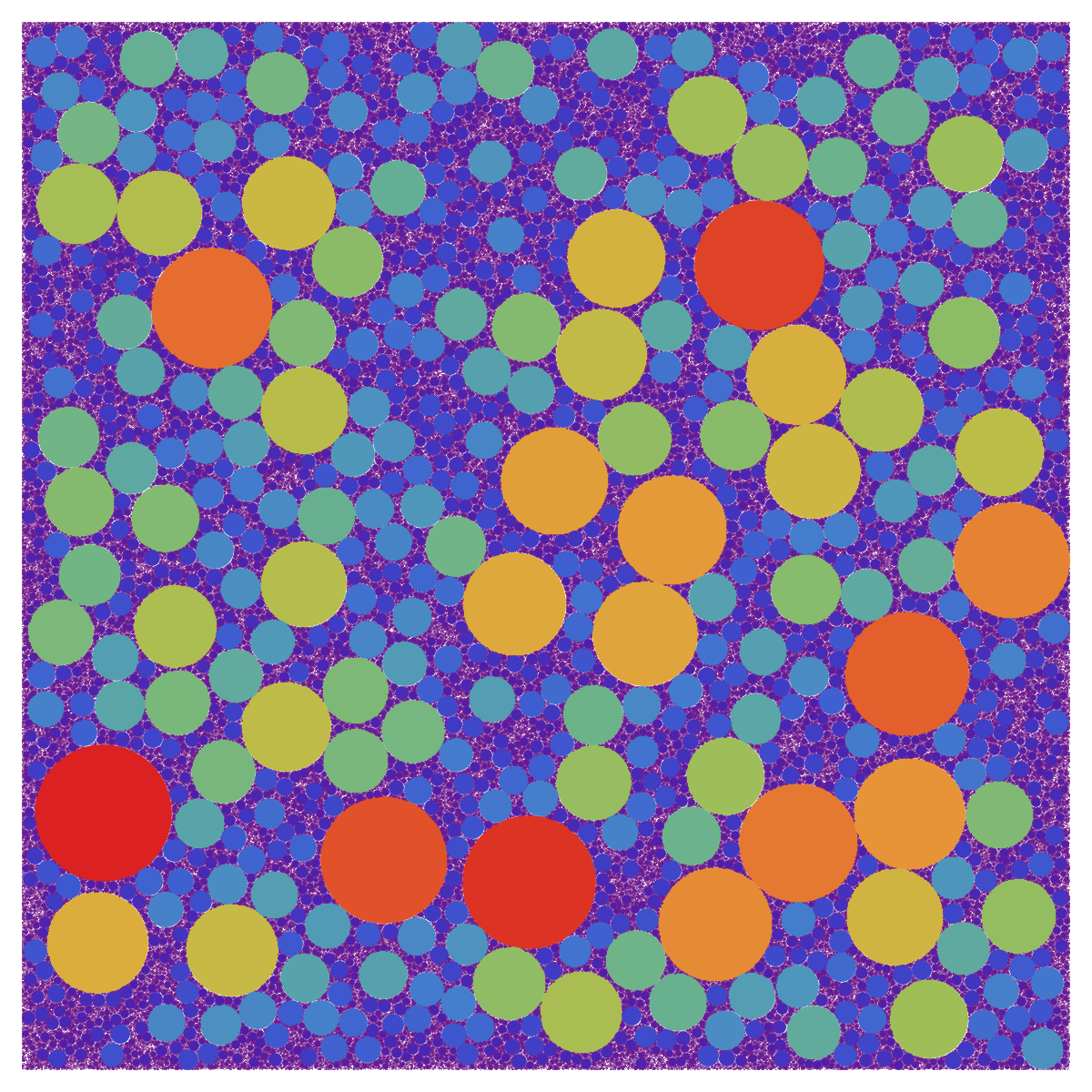}
\caption{(Top) A grid of $5 \times 5$, where each cell contains $M = 5000$ disks placed with the RSA algorithm described in Ref.~\cite{cherny23}. The total number of the disks $N=125000$. (Bottom) A set of $N = 125000$ disks arranged with the DT algorithm. In both cases, the disks are compactly packed and randomly distributed, and their radii follow power-law distributions with the exponent $D=1.5$. Disks of the same radius are displayed in the same color. \label{fig1}}
\end{figure}

\subsection{Delaunay triangulation algorithm}
\label{sec:DTalg}

For the DT algorithm, we use a smoother implementation of the same power-law distribution as for the RSA grid:
\begin{equation}
    r_i = R \left(\frac{m}{m+i-1}\right)^{1/D},\  \textrm{for}\  i = 1, ..., N,
    \label{distr}
\end{equation}
where $m=k^2$. When $N\gg m$, the smallest radius $a = R \big[m/(N+m-1)\big]^{1/D}$ only slightly deviates from $a = R ({m}/{N})^{1/D}$, which is realized in the RSA grid (Sec.~\ref{RSA}). The largest radius $R$ should be chosen to ensure the densest packing in the thermodynamic limit. It is achieved when the area of the embedding square equals the total area of the disks:
\begin{equation}
s^2/m= \pi R^2m^{2/D-1}\sum_{i=m}^{\infty}i^{-2/D}=\pi R^2m^{2/D-1}\zeta (2/D,m),
    \label{area}
\end{equation}
where $s$ is the edge length of the square, and $\zeta (t,a)=\sum _{k=0}^{\infty } (a+k)^{-t}$ is the Hurwitz zeta function. In the thermodynamic limit $m\to\infty$ and $N/m = \text{const}$, Eq.~(\ref{area}) yields $s^2/m\to\pi R^2 D/(2-D)=\text{const}$ in full analogy with Sec.~\ref{RSA} at $m=k^2$. Note that the limit $\lim_{m\to\infty}s^2/m=\pi R^2 D/(2-D)$ is also obtained with the continuous power-law distribution as it should be.

\begin{figure}[tb]
\includegraphics[width=\columnwidth, trim={1cm 1cm 1cm 1cm}]{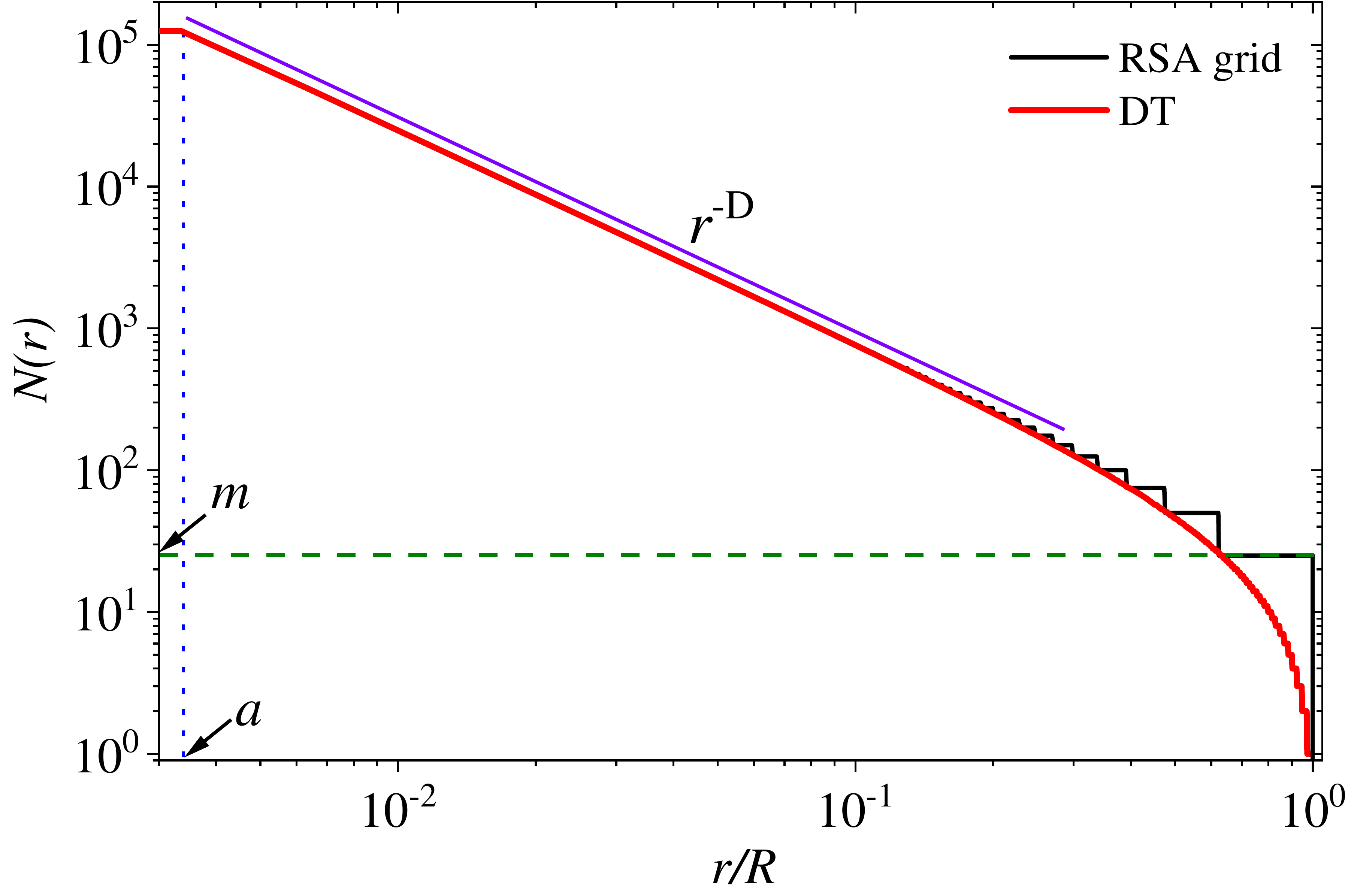}
\caption{Cumulative power-law distribution in units of the largest radius $R$ for the two structures shown in Fig.~\ref{fig1}. The vertical dashed line (blue) denotes the minimum radius $a$. The horizontal dashed line (green) indicates the number of cells in the RSA grid.
\label{fig2}}
\end{figure}

Figure \ref{fig2} compares the distributions of radii used in the RSA grid and those given by Eq.~(\ref{distr}). We observe that they both decay with the power-law $r^{-D}$ and are indistinguishable in practice, although the latter distribution is smoother.

To enable a compact packing of the disks within a finite area, the exponent of the distribution should be restricted \cite{cherny23} to $D < 2$.
Disks are arranged into a square by placing them from largest to smallest. As the set of already placed disks grows, the computational costs of checking for their collisions increase. Thus, we use the DT to select neighbours of a newly placed disk and thus reduce the number of required calculations. The DT covers a convex hull of a given set of points by a set of triangles in such a way that no point lies inside the circumcircle of any triangle. There are very effective implementations of the triangulation that need only $O(\log N)$ operations for both adding a new point in the triangulation and finding neighbours of the point. We employ the code from the CGAL software package \cite{cgal}.

The algorithm consists of the following steps:
\begin{enumerate}
    \item Initially, positions of the first six disks are randomly determined across the entire square.

    \item To place a new disk, a random point is chosen for its center. The function \linebreak \emph{vertices\_on\_conflict\_zone\_boundary} from the CGAL package identifies a subset of already placed disks that are neighbours to the new disk and most likely to intersect with it. If the randomly selected point is too close to their neighbours and collisions appear, the algorithm searches for an empty space inside the polygon whose vertices are the neighbours of the selected point. To find an empty space, the new disk is bounced back with the shift vector that amounts to the sum of vectors of all found overlappings. If no collisions occur then we go to step 3. If a collision with the neighbours still exists, a new random point is chosen once more and step 2 is repeated again. If the maximum number of unsuccessful attempts is reached, the algorithm is restarted from step 1.

    \item Before placing a new disk, the last check for collisions with all previously placed disks is performed. This is because the selection of neighbors with the DT might not work perfectly, although the probability of failure is very small.

    \item The center of the newly placed disk is added to the triangulation set and the algorithm proceeds from step 2.
\end{enumerate}

This algorithm does not place disks  completely randomly. Although each new disk initially drops at a random point, it then simulates repulsion from its neighbors and the boundaries of the square. For $m=1$, the first disk is placed at a corner of the square; for $m > 1$ it is positioned at a random point. The larger the number of disks and the packing fraction, the more efficient this algorithm is in comparison to a simple random placement of the disks such as RSA. The DT algorithm reduces the search for an empty space to the vicinity of each added disk, which makes it especially effective at high packing fractions. The performance increase is approximately proportional to $1/(1-F)$, where $F$ is the current filling rate.

Figure \ref{fig1} (Bottom) shows a trial with $N = 125000$ disks generated by the DT-based algorithm with distribution (\ref{distr}) at the same control parameters as described in Sec.~\ref{RSA}. The color coding is the same as in Fig.~\ref{fig1} (Top). In the RSA grid, the spatial distribution of disks is more homogeneous by its construction. However, the spatial distributions of RSA and DT are almost indistinguishable on scales of order $r\lesssim s/k$.

\section{The structure factor}
\label{sec:sq}

We define the structure factor through the density fluctuations over the background of the density mean value such that
\begin{equation}
S(q)\equiv\langle\delta\rho_{\bm{q}}\delta\rho_{\bm{-q}}\rangle_{\hat{q}}/N,
\label{sq}
\end{equation}
where the Fourier components of the density fluctuations are given by
\begin{align}\label{deldens}
 \delta\rho_{\bm{q}}=\sum_{j=1}e^{-i\bm{q}\cdot\bm{r}_{j}}-N F_{\mathrm{sq}}(\bm{q}s)
\end{align}
with
 $F_{\mathrm{sq}}(\bm{q}s) =\int d\bm{r}e^{-i\bm{q}\cdot\bm{r}}/{s^2}=\frac{\sin q_x s/2}{q_x s/2}\frac{\sin q_y s/2}{q_y s/2}$
being the the scattering amplitude of the embedding square normalized to one at $q=0$. The origin of coordinates is chosen in the center of the
square. The brackets $\langle\cdots\rangle_{\hat{q}}$ stand for the average over all directions of unit vector  $\hat{q}$ along $\bm{q}$.
Note that in the previous papers \cite{cherny22,cherny23}, the structure factor was defined without subtracting the scattering amplitude of the background. Such a definition leads to the appearance of a spike of order of $N$ in the range $q\lesssim 2\pi/s$, which is not convenient to study the thermodynamic limit. The definition (\ref{sq}) enables us to effectively eliminate the spike.

\begin{figure}[tb]
\includegraphics[width=0.95\columnwidth]{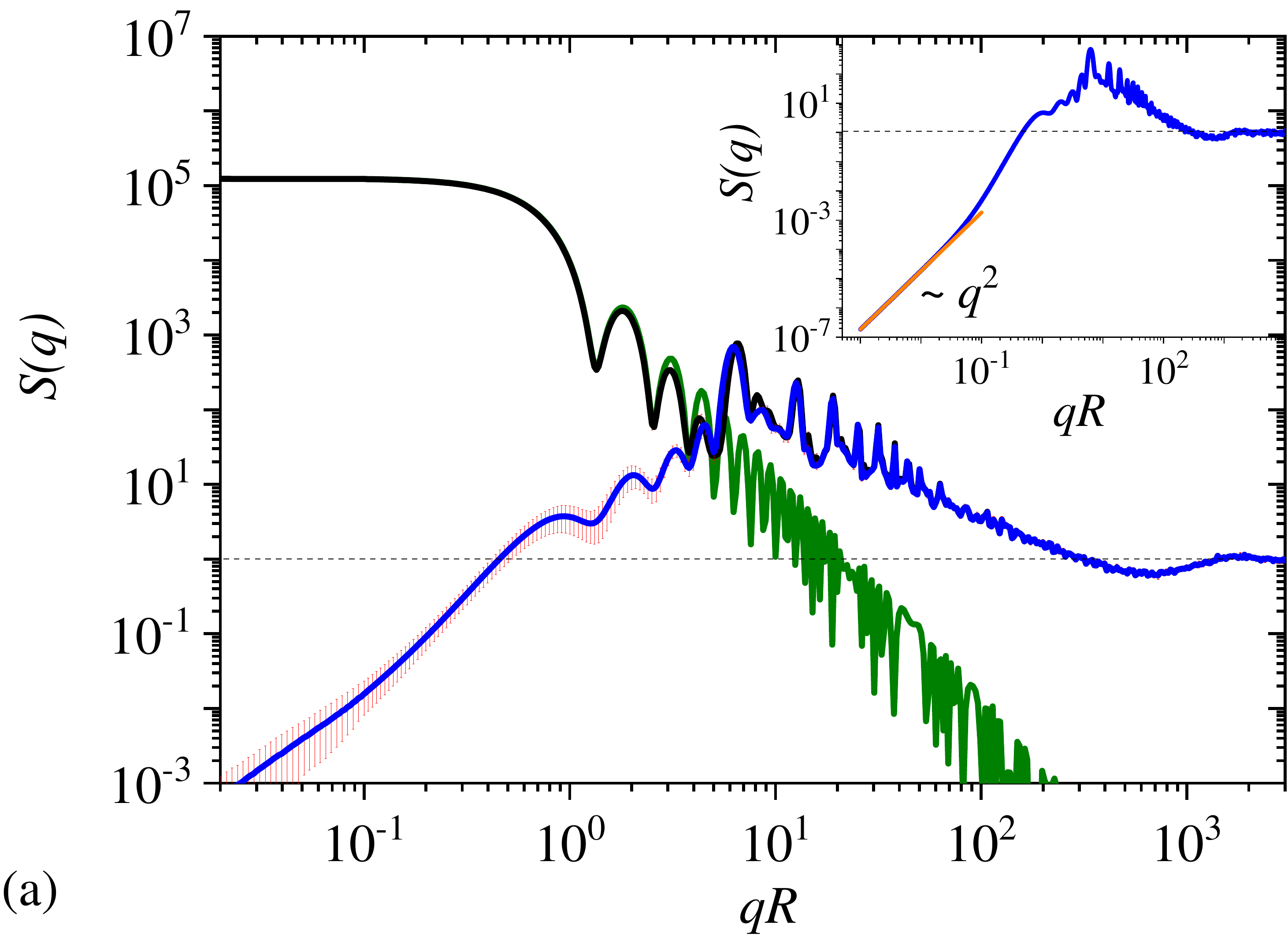}
\includegraphics[width=0.95\columnwidth]{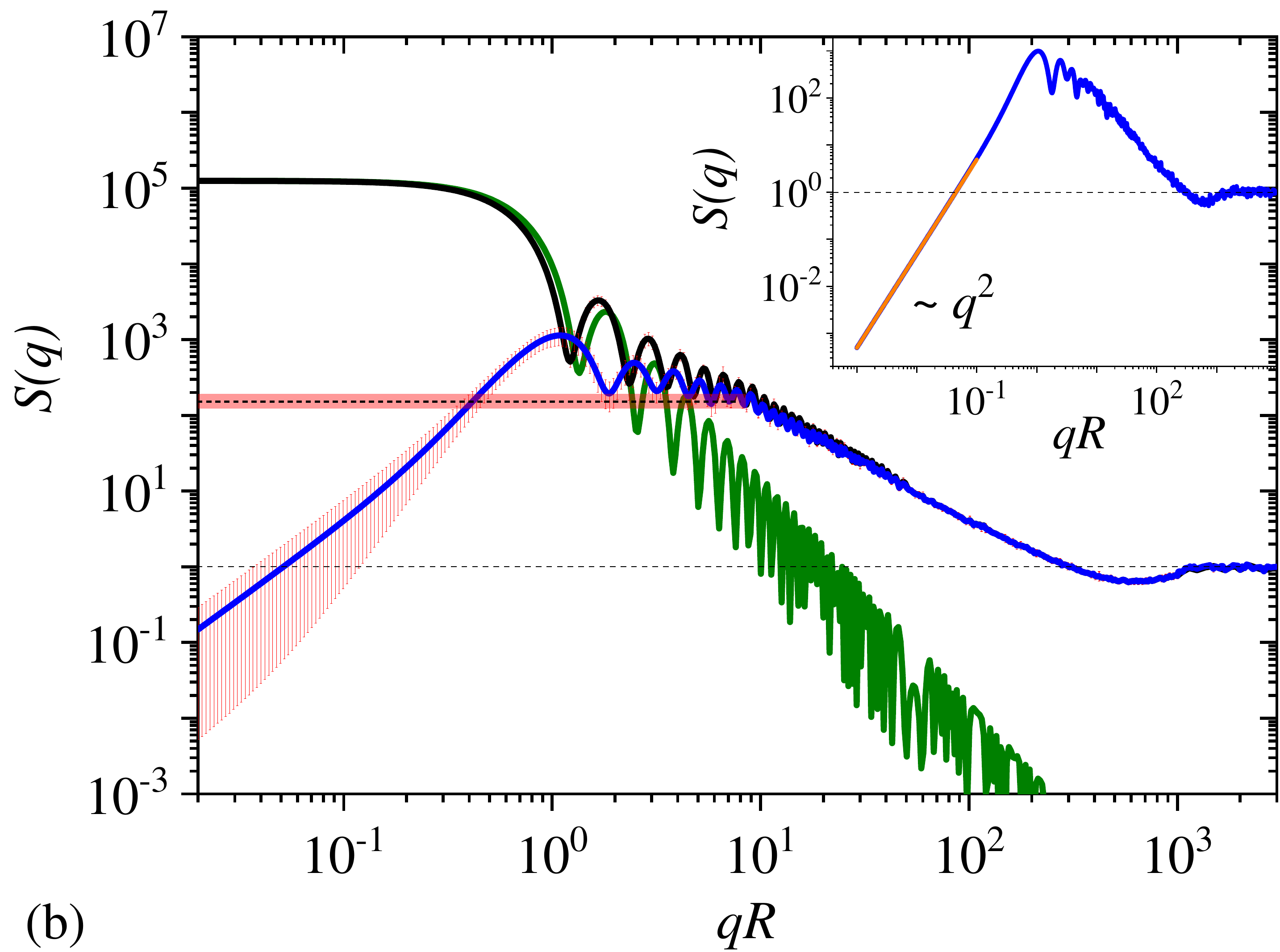}
\caption{\label{fig3} The structure factor \eqref{sq} (blue) corresponding to the density fluctuations \eqref{deldens} in units of the inverse radius $1/R$ of the largest disk. The structure factor \eqref{sq} (black) corresponding to the densities, which are given by \eqref{deldens} without the second term. The scattering intensity from the embedding square $N \langle F^2_{\mathrm{sq}}(\bm{q}s)\rangle_{\hat{q}}$ is shown in green. (a) RSA grid. (b) DT. In both cases, the red bars represent errors (the standard deviations for 20 trails). The number of disks $N = 125000$ and $s/R \simeq 15$. The corresponding structures are shown in Fig.~\ref{fig1}. The inset figures represent the exact agreement between a single trial (blue) and the analytical relation \eqref{sqsmall} (orange) at small values of the wave vector $q\lesssim 2\pi/s$.
}
\end{figure}

Figure \ref{fig3} shows the structure factor (\ref{sq}) for the dense packings depicted in Fig.~\ref{fig1}. Each curve is calculated numerically with 20 trials. We observe that the errors are almost negligible starting from the lower border of the fractal range $q \gtrsim 2\pi/R$, see the discussion in Sec.~\ref{sec:FR} below. This suggests that for $q \gtrsim 2\pi/R$ the structure factor is independent of a specific packing configuration in the limit of high packing fraction. However, when $q\lesssim 2\pi/R$ the errors are quite pronounced, especially for the DT-based curves.

To illustrate graphically how the structure factor is obtained, Fig.~\ref{fig3} shows also the structure factor given by Eq.~\eqref{sq}  but without subtracting the scattering amplitude of the background square in Eq.~\eqref{deldens} (black curve). We observe that that for $q\lesssim 2\pi/s$ the dominant contribution comes from the embedding square (green curve), while for $q \gtrsim 2\pi/R$, the main contribution arises from the structure factor of a single cell. The crossover point lies somewhere in between. The contribution of a single cell was s studied in detail in the previous paper \cite{cherny23}.

\begin{figure}[tb]
\includegraphics[width=.95\columnwidth]{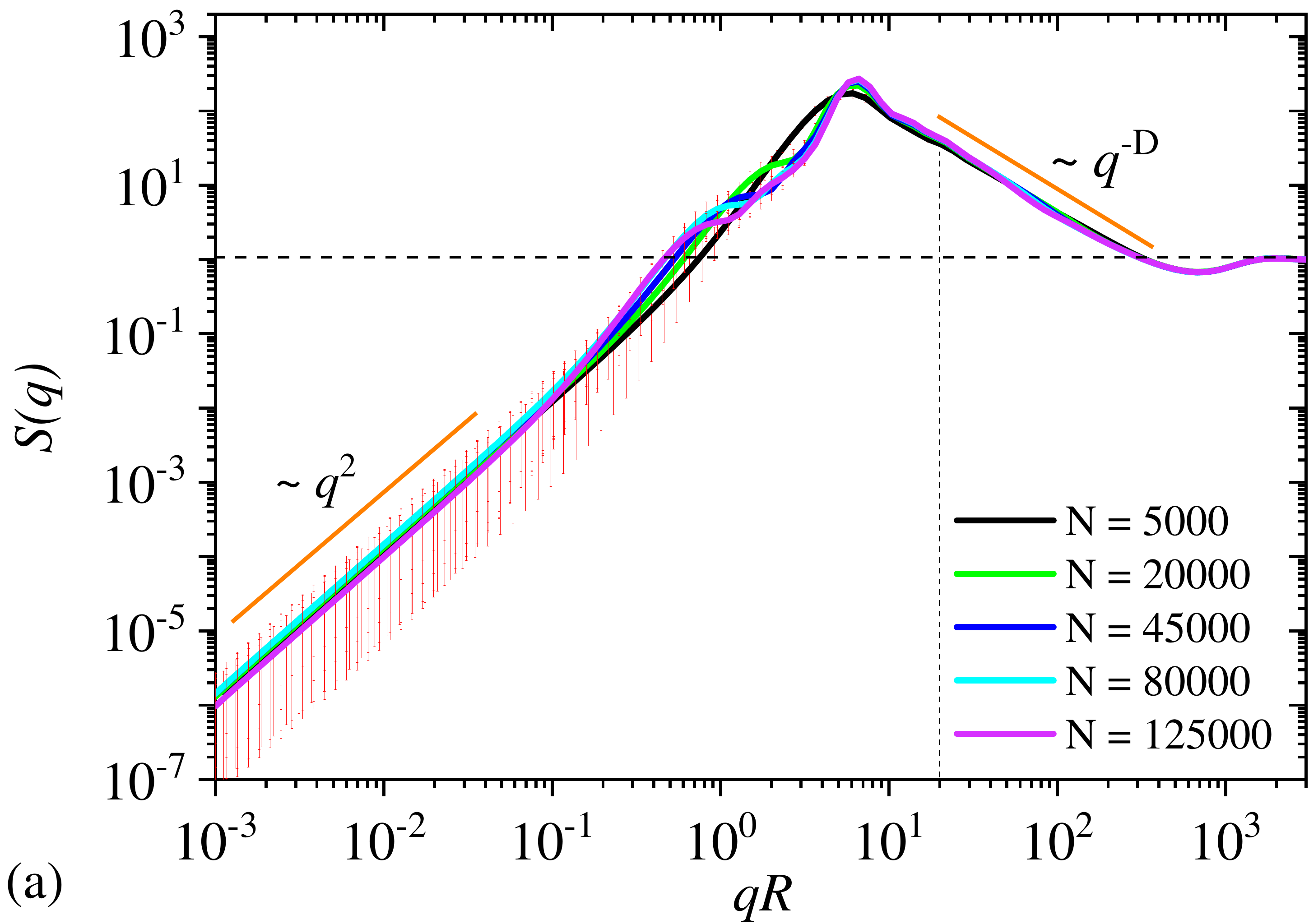}
\includegraphics[width=.95\columnwidth]{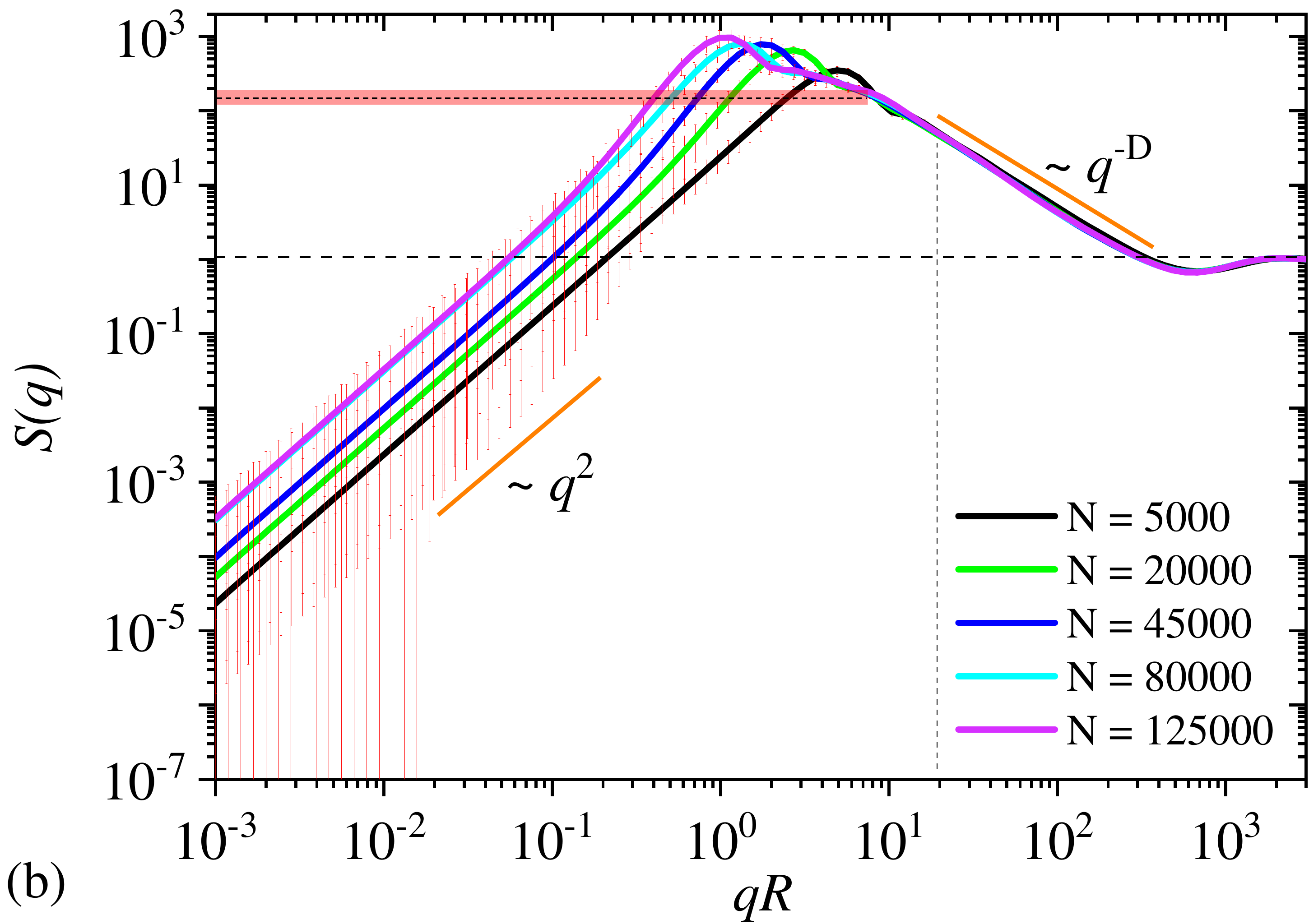}
\caption{The smeared structure factor (\ref{sq}) vs the wave vector (in units of the inverse radius $1/R$ of the largest disk) at various numbers of particles $N$. (a) RSA grid. (b) DT. In both cases, the red bars represent the errors. The vertical dotted lines indicate the start of the fractal range. The horizontal short-dashed line in panel (b) at $148 \pm 26$ shows the structure factor $S(0)$ obtained from Eq.~(\ref{fluct}), which effectively excludes the finite-size effects (see the discussion in Sec.~\ref{S0TD}). \label{fig4}}
\end{figure}

A quantitative analysis of the structure factor is carried out by smearing the curves in Fig.~\ref{fig3} with a log-normal distribution function, as described in Ref.~\cite{cherny23}. This procedure eliminates the maxima and minima by smoothing them and provides a reliable way to estimates the power-law exponents by a linear fit in a double-logarithmic plot. Moreover, smoothing is a standard procedure in analyzing small-angle scattering data \cite{wignall91}.
The results of smearing are shown in Fig.~\ref{fig4} for various number of disks $N$. For the RSA grid, the chosen values of $N$ correspond to different grid sizes ($1\times1$ at $N = 5000$, $2\times 2$ at $N = 20000$ etc). In particular, in Fig.~\ref{fig4} the violet curve at $N = 125000$ is the smeared version of the blue curve in Fig.~\ref{fig3}.

\subsection{Small momenta}
\label{smallQ}

\subsubsection{Finite-size effects}

Upon expanding Eq.~(\ref{deldens}) in powers of momentum, we obtain  $\delta\rho_{\bm{q}} = -i N\bm{q}\cdot\bm{r}_\text{c}+O(q^2)$ with $\bm{r}_\text{c}= \frac{1}{N}\sum_{j=1}^{N}\bm{r}_j$ being the center-of-mass position of the entire set of points. After substituting this formula into the definition (\ref{sq}) and (\ref{deldens}) and using the identity $\langle(\bm{q}\cdot\bm{r}_\text{c})^2\rangle_{\hat{q}}=q^2r_\text{c}^{2}/2$ in 2D, we are left with the following structure factor for $q\lesssim 2\pi/s$
\begin{align}
    S(q)\simeq\frac{N}{2}q^2r_\text{c}^{2}.
    \label{sqsmall}
\end{align}

The vector $\bm{r}_\text{c}$ is nonzero solely due to density fluctuations. When a distribution of the points $\bm{r}_{j}$ over the square is random and uniform, the variance $\left\langle\left(\sum_{j=1}^{N}\bm{r}_j\right)\cdot\left(\sum_{j=1}^{N}\bm{r}_j\right)\right\rangle$ is proportional to $N$. It follows that the structure factor (\ref{sqsmall}) is independent of $N$ in the thermodynamic limit as it should be. Nevertheless, in this limit, the range $q\lesssim 2\pi/s$ shrinks to zero, and, hence, formula (\ref{sqsmall}) essentially characterizes a
finite-size effect.

Figure \ref{fig4} illustrates these features. For the RSA grid, the structure factor remains practically unchanged with increasing $N$ when $q\lesssim 2\pi/s$. For the DT based algorithm, the structure factor approaches a saturation curve within this region as $N$ increases. For $N > 80000$, the corresponding scattering curves are practically indistinguishable. Note that all the curves fall within the error bars when $q\lesssim 2\pi/s$.

Figure \ref{fig4} shows that the power-law exponent at small momenta takes the value 2 in accordance with Eq.~\eqref{sqsmall}. Moreover, the coefficient of $q^{2}$ in Eq.~\eqref{sqsmall} is exactly recovered from the corresponding curves (see the insets of Fig.~\ref{fig3}).

\subsubsection{Thermodynamic limit and hyperuniformity}
\label{S0TD}

We emphasize that Eq.~(\ref{sqsmall}) describes a finite-size effect, which is valid for an \emph{arbitrary} distribution of points over the square, and is not related to the hyperuniformity~\cite{torquato2003,torquato2018} of the system. We recall that hyperuniformity, an important concept in condensed matter physics for the classification and structural characterization of crystals, quasicrystals and special disordered point configurations~\cite{torquato2018}, assumes that $S(q) \rightarrow 0$ as $q \rightarrow 0$ \textit{after} the thermodynamic limit is reached.




If a many-particle system is not hyperuniform then no plateau in a double-log diagram should be visible in the structure factor (\ref{sq}), (\ref{deldens}) for small momenta $2\pi/s \ll q\ll 2\pi/\xi$, where $s$ and $\xi$ is the total size of the system and its characteristic correlation length, respectively. The latter appears, say, due to interactions between the particles. In the thermodynamic limit, the total size of the system tends to infinity, and  the  plateau can be treated as the limiting value $S(q=0)$. We indeed observe a reminiscence of the plateau in Figs.~\ref{fig3}(b) and \ref{fig4}(b) when $2\pi/s \ll q\lesssim 2\pi/R$. Unfortunately, the used ratio $s/R\simeq 15$ is not big enough for direct and reliable numerical evidence of the plateau presence.

Nevertheless, one can obtain $S(0)$ by calculating the fluctuations of the number of points inside an imaginary circle such that \cite{torquato2003}
\begin{align}
\sigma^2_n(\rho) = \langle n^2(\rho) \rangle - \langle n(\rho) \rangle^2\simeq S(0)\langle n(\rho) \rangle,
    \label{fluct}
\end{align}
where $n(\rho) = \sum_{j=1}^N \Theta (\rho - |\bm{r}_j - \bm{x}_i|)$ is is the number of points inside a circle of radius $\rho$, centered at some position $\bm{x}_i$ inside the square. The average is taken over randomly chosen centers $\bm{x}_i$.  An ensemble of two thousand centers was used in our numerical simulations. Equation \eqref{fluct} is satisfied when $\xi\ll \rho \ll s$. The first equality ensures that $\langle n(\rho) \rangle$ is proportional to $\rho^2$, while the last inequality is needed to exclude the boundary effects.  It follows from Eq.~(\ref{fluct}) that $S(0) \geqslant 0$. After the thermodynamic limit, one can consider the long-range limit $\rho\to\infty$; then Eq.~\eqref{fluct}, describing the fluctuations of particles in the grand canonical ensemble, becomes exact.

If $S(0)=0$, a many-body system is hyperuniform  \cite{torquato2018}. In our case, $\sigma^2_n(\rho)/\rho^2 \simeq \text{const}\not=0$ when $R\lesssim\rho\lesssim s$, so hyperuniformity is not observed. The linear fit in this range yields $S(0)=\sigma^2_n(\rho) / \langle n(\rho) \rangle = 148 \pm 26$, which tells us that our system is \textit{not} hyperuniform. The value of $S(0)$ is shown in Figs.~\ref{fig3}(b) and \ref{fig4}(b) as the horizontal dashed line. The ``plateau" in the structure factor near $2\pi/R$  and the horizontal line are of the same order but do not coincide. However, in the thermodynamic limit $s/R\to\infty$, $S(0)$ obtained using Eq.~(\ref{fluct}) and the plateau in the structure factor (\ref{sq}) should merge perfectly.

These findings are consistent with other numerical and experimental studies, which have also reported non-hyperuniformity in polydisperse systems~\cite{xu10,kurita10}. However, some papers reported the existence of hyperuniformity  in  polydisperse jammed packings of spheres \cite{berthier11}. In these papers, hyperuniformity is defined through fluctuations of the local \emph{volume} fraction but not the number of \emph{points}.
In the context of jammed \textit{monodisperse} spheres, the situation appears more definitive, with evidence suggesting that these systems are either hyperuniform~\cite{donev05} or 'nearly hyperuniform'~\cite{ikeda15}.

\subsection{High momenta: Fractal region and beyond}
\label{sec:FR}

The fractal range in reciprocal space is given by \cite{cherny23}
\begin{equation}
2\pi / R \lesssim q \ll 2\pi/a.
    \label{fr}
\end{equation}
The lower and upper borders of the fractal region are related to the largest and smallest radii, respectively, in the studied configuration. The left border of this range is indicated by a vertical dotted line in Figs.~\ref{fig4} and \ref{fig5}. Within the fractal range, the structure factor decays as $q^{-\alpha}$. A linear fit in the range $22 \lesssim qR \lesssim 131$ gives the power-law exponent $\alpha = 1.508 \pm 0.009$. This confirms numerically that $\alpha=D$.

Beyond the fractal region, when $q \gtrsim 2\pi/a$, the structure factor attains its asymptotics $S(q) \simeq 1$, which is indicated by the black horizontal dotted line in Figs.~\ref{fig3}--\ref{fig6}. The lengthy region where the structure factor lies below one appears due to finite values of smallest radius $a$ (see the detailed discussion in Ref.~\cite{cherny23}).

\subsection{Random-sequential-addition grid \textit{vs} Delaunay triangulation algorithm}

\begin{figure}[tb]
\includegraphics[width=0.95\columnwidth]{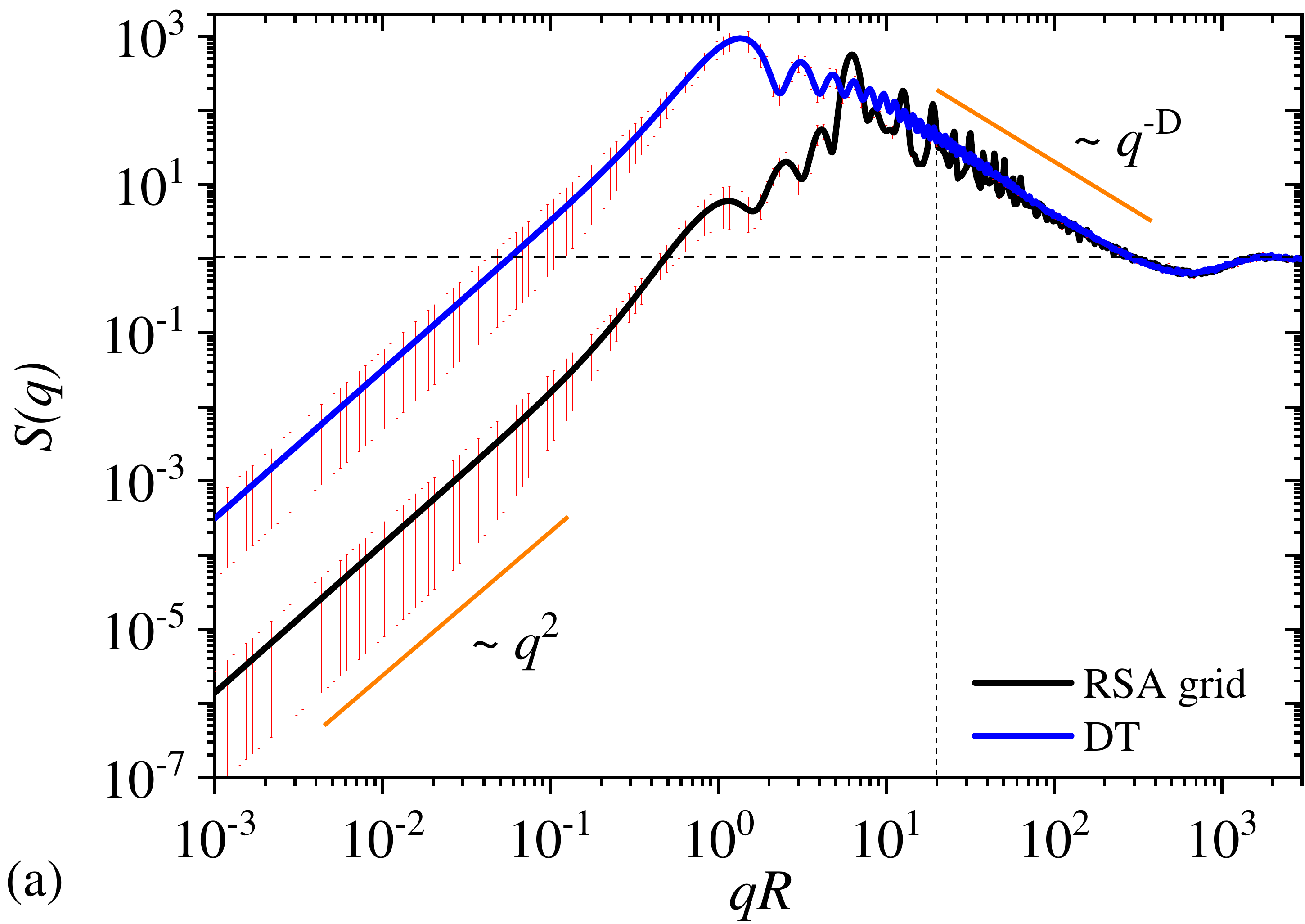}
\includegraphics[width=0.95\columnwidth]{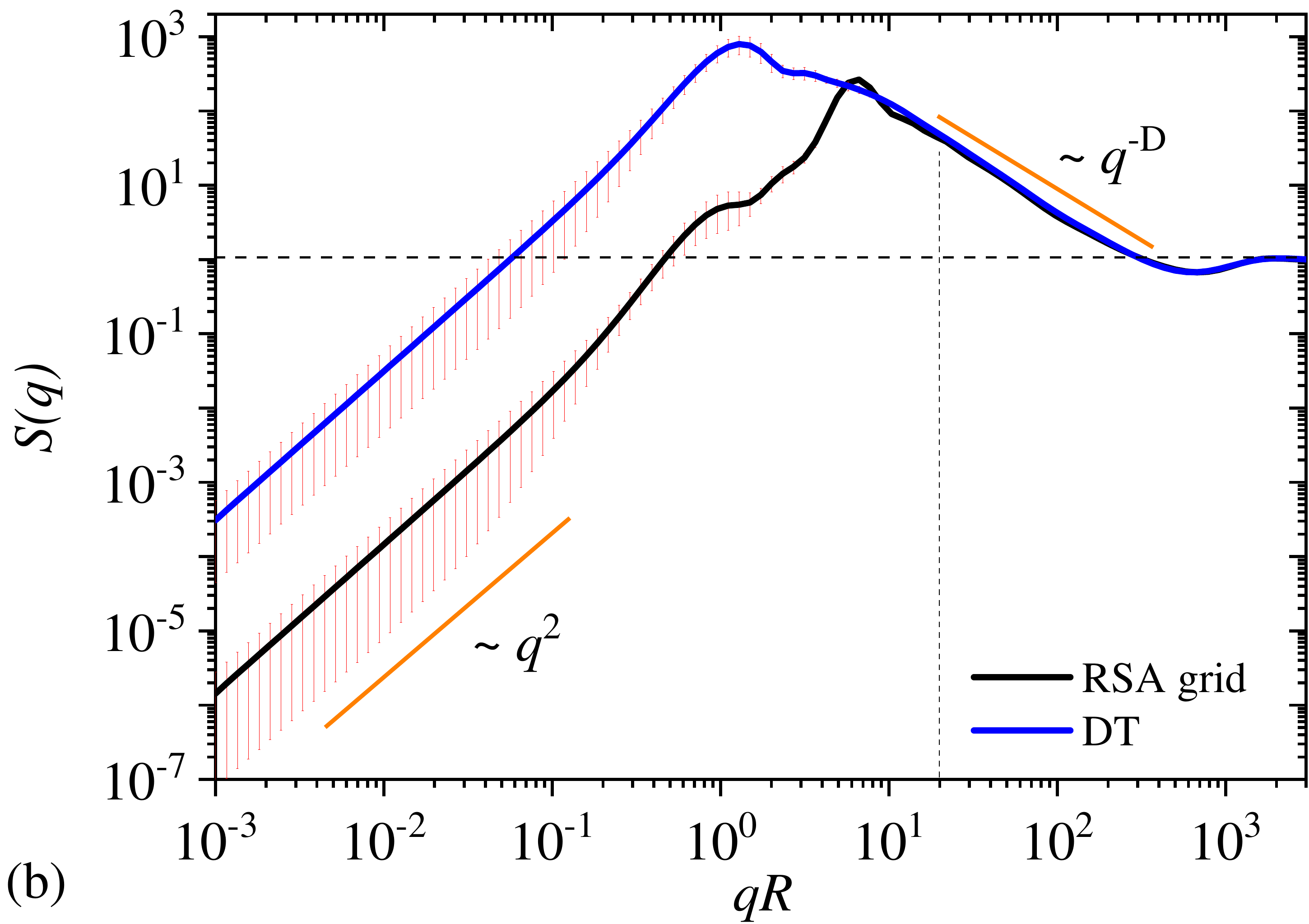}
\caption{(a) The structure factors (\ref{sq}) vs the wave vector (in units of the inverse radius $1/R$ of the largest disk) for the RSA grid and DT at $N = 80000$. (b) The corresponding smeared structure factors.
\label{fig5}}
\end{figure}

In Fig.~\ref{fig5}, a comparison is given between the structure factors (\ref{sq}) calculated with the RSA grid and DT-based algorithms. The results show an excellent agreement between the curves starting from the lower border of the fractal range.  For small momenta $q\lesssim2\pi/R$, the curves are different, because the spatial long-range correlations of the disk positions are artificially changed at the distances bigger than the cell size, which is of the order of $R$ (see the discussion in Sec.~\ref{RSA}). Nevertheless, the RSA grid and DT packing algorithms produce the same fractal exponent, which coincides with the power-law exponent of the size distribution.

For small momenta $q \gtrsim 2\pi/a$, the corresponding coefficients of $q^{2}$ in Eq.~\eqref{sqsmall} differ by two orders of magnitude or so, see Fig.~\ref{fig5}. This is related to the fluctuations of the center-of-mass position of the disk centers. The fluctuations are artificially suppressed in the RSA grid by its construction (see Sec.~\ref{RSA}) and as a consequence, the coefficient is significantly reduced.

To analyze the behavior of the structure factor in the intermediate range between the small-momenta regime of $q$ and the fractal regions, the DT method is more preferable because of its efficiency. This approach effectively reduces the artifacts typically observed with the RSA grid method (see also the comparison between RSA and DT methods in Sec. III D below).

Within the context of efficient packings, other protocols have been developed. For instance, the centroid/quantizer approach, as described by Klatt et al.~\cite{Klatt19}, has been developed for generating packings in two and three dimensions.  The method is based on Lloyd's iterative algorithm that converges to centroidal Voronoi tessellations, starting from a given initial point set. These packings exhibit an anomalous suppression of long-wavelength density fluctuations and thus are ``effectively hyperuniform". However, it is uncertain whether this method is applicable to packing of disks with a given power-law distribution of radii.


\subsection{Dependence of the structure factor on density}
\label{sec:densities}

In the previous sections, the number of particles was varied but the packing fraction and  density $N/s^2$ were kept constant. In this case, the length of the fractal range and the power-law exponent in the structure factor remain intact (see Sec.~\ref{sec:FR} and Fig.~\ref{fig4}).

\begin{figure}[tb]
\includegraphics[width=0.95\columnwidth]{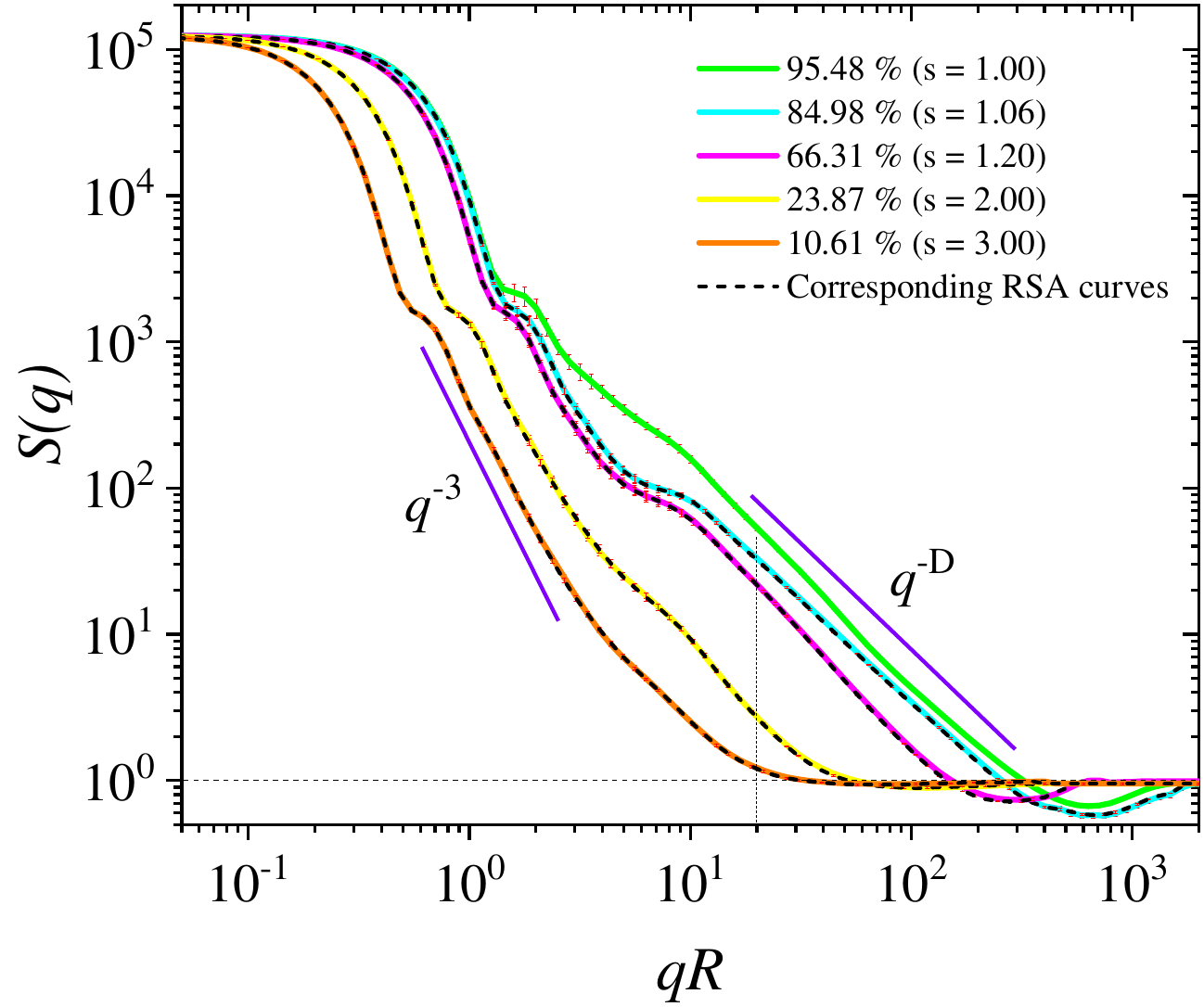}
\caption{Comparison of the smeared DT structure factor (continuous lines) and the corresponding RSA structure factor (black dashed lines) as a function of the wave vector (in units of the inverse radius $1/R$ of the largest disk) for various packing fractions. The structure factor is given by  Eq.~\eqref{sq} and \eqref{deldens} without the second term. The total number of disks is fixed and equal to $N = 125000$. The vertical dotted line marks $q=2\pi/R$, which is of the order of the lower border of the fractal range.\label{fig6}}
\end{figure}

Then, the question arises what happens when the \textit{density} decreases? To elucidate the relationship between the structure factor and density variations, we generate random configurations of disks at fixed $N$ and various linear system sizes $s$. The DT and RSA algorithms were used across a range of packing fractions. We emphasize that the RSA algorithm was employed to the \emph{entire} system for all the packing fractions except the maximum fraction of $95.48~\%$ because of its substantial computational requirements. The structure factor (Fig.~\ref{fig6}) for each packing fraction is determined by averaging the outcomes of 10 distinct trials. We observe that both algorithms yield the same results across all evaluated momenta and packing fractions.

\begin{table}
    \caption{\label{tab:alpha} The power-law exponents for the structure factors shown in Fig.~\ref{fig6}. The exponents are obtained by linear fits within the fractal ranges $R q_{\text{min}}~\lesssim R q~\lesssim R q_{\text{max}}$. }
    \begin{ruledtabular}
    \begin{tabular}{ccccc}
       \,\, packing fraction \,\,&\,\, exponent \,\,$\,\,\alpha$ \,\,&\,\, $R\,q_{\text{min}}$ \,\,&\,\, $R\,q_{\text{max}}$ \,\,\\
         95.48\% & 1.52 $\pm$ 0.02 & 20 & 200 \\
         84.98\% & 1.40 $\pm$ 0.01 & 20 & 160\\ 
         66.31\% & 1.56 $\pm$ 0.01 & 20 & 153\\ 
         23.87\% & 1.49 $\pm$ 0.05 & 16 & 27\\
         10.61\% & 1.43 $\pm$ 0.04 & 8 & 11\\
    \end{tabular}
\end{ruledtabular}

\end{table}

The power-law exponent $\alpha$ remains approximately equal to $1.5$ despite the decrease in packing fraction, although the extent of fractal range diminishes. An impact on the power-law exponent from log-normal polydispersity could emerge due to the limited fractal range. To eliminate this potential impact, we perform linear fitting directly on the monodisperse structure factor. The results are presented in Table~\ref{tab:alpha}.

Despite the variations of $\alpha$ around $1.5$, we do not reach the values of about 1.28 reported in Ref.~\cite{monti23}. This is in contrast to jammed packings generated by discrete element method simulations: it has been reported \cite{monti23} that the structure-factor exponent is independent of the exponent of the power-law size distribution in three dimensions and only slightly depends on it in two dimensions.  The authors of the paper \cite{monti23} attributed these discrepancies to the packing protocols used, which could potentially differentiate between the final particle structures. However, as shown above, RSA, RSA grid, and DT protocols all give the same power-law exponent, suggesting that in general, in the limit of very high packing, all protocols generating dense random packings give the same exponent. This is still an open question and additional investigations are needed. Note that the resulting value of $\alpha$ is quite sensitive to the choice of borders of the fractal range, particularly of the upper border, which is followed by a pronounced change of the curve towards its asymptotic value.

At low packing fractions, the system becomes dilute, and thus the spatial correlations decay substantially. As a consequence, the distribution of disks over the square becomes more uniform. Then the scattering pattern more closely resembles the structure factor of the embedding square as one can see in Fig.~\ref{fig6}. In particular, the Porod power-law $q^{-3}$ is observed.

\section{Conclusions}
\label{sec:concl}

In this work, we examine the structure factor $S(q)$ (\ref{sq}) to study the correlation properties of densely packed disks with a power-law size distribution in the thermodynamic limit. The thermodynamic limit is simulated with different methods of packings: the RSA grid and a more effective algorithm developed in this paper, which is based on the DT, see Sec.~\ref{sec:DTalg}.

The spike in the structure factor at small momenta, which is proportional to the total number of the disks, is excluded by subtracting the scattering amplitude of the embedding square (see Sec.~\ref{sec:sq}). Apart from formal reasons, this definition can be quite appropriate from the physical point of view, particularly for studies of internal structures using small-angle neutron scattering with the contrast variation method, see e.g. Ref.~\cite{krueger22}.

The main result of the paper is that in the thermodynamic limit, the fractal range in $S(q)$ is still observable for both algorithms: its borders (\ref{fr}) are determined by ranges of the size distribution, and the fractal exponent coincides with the exponent of the power-law size distribution \cite{cherny23}.

We investigate the dependence of structure factor on density, see Sec.~\ref{sec:densities}. The decrease in packing fraction leads to narrowing of the fractal range, while the power-law exponent $\alpha$ in the structure factor remains approximately equal to the exponent $D$ of the size distribution.

The finite size effects are studied as well. It is shown (see Sec.~\ref{smallQ}) that in the range of small momenta, $S(q) \propto q^{2}$ and it is independent of the number of particles, although this range shrinks in the thermodynamic limit.  We derive the expression for the coefficient of $q^2$, which is related to the fluctuations of the center-of-mass of the disk  positions.

The obtained results can be used for analyzing experimental small-angle scattering data from densely packed objects, whose centers have a much higher scattering length density as compared to its surrounding shell.

\acknowledgments

The authors acknowledge support from the JINR--IFIN-HH projects.

\appendix

\section{RSA complexity}
\label{sec:RSAcompl}

\begin{figure}[!htb]
\includegraphics[width=0.75\columnwidth]{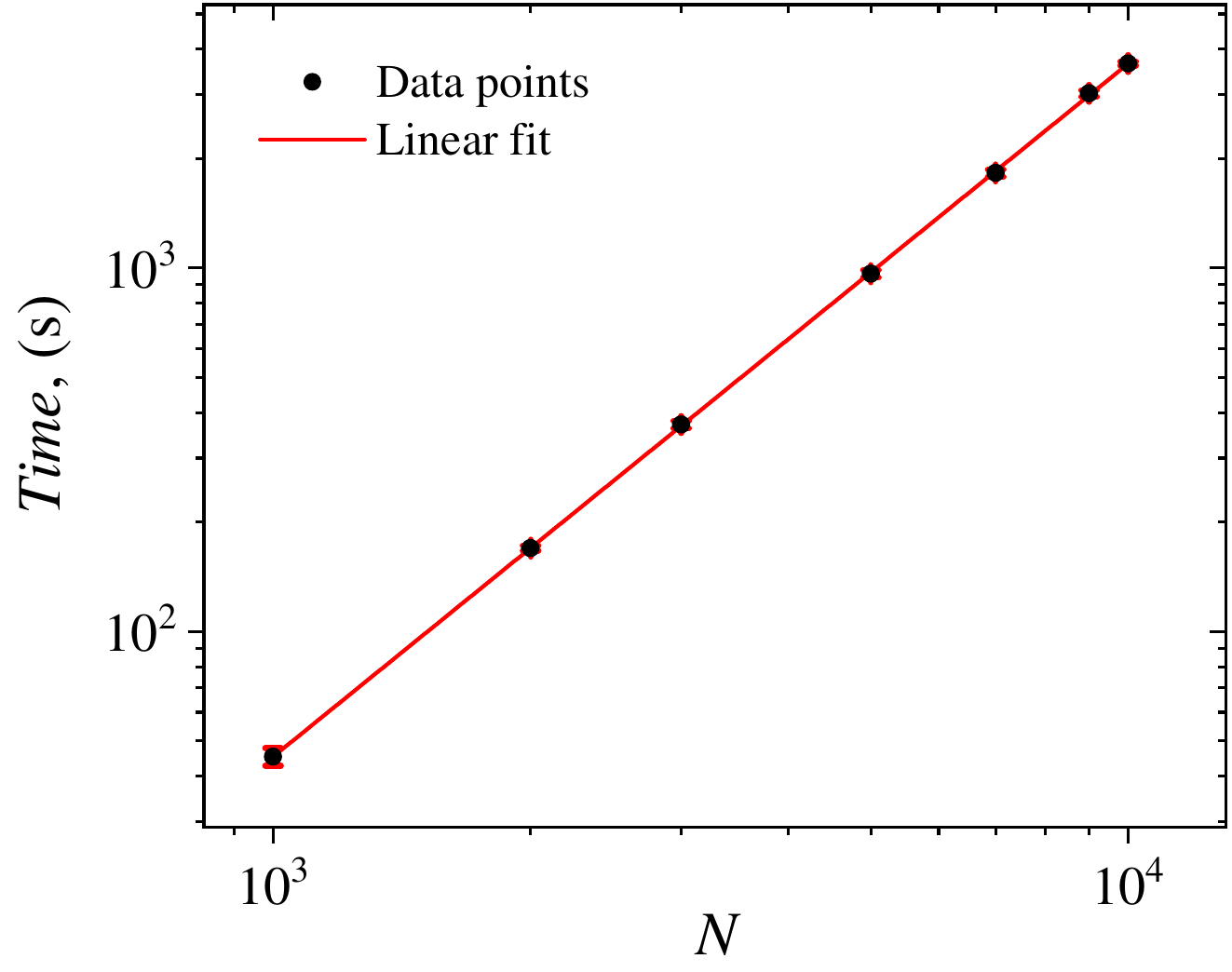}
\caption{Variation of the computation time required for the RSA algorithm to generate a dense pattern of $N$ disks (black dots) along with a linear fit (red continuous line).
\label{appendix}}
\end{figure}

To investigate the time complexity of the RSA algorithm, we recorded the computation time required to generate dense packing patterns at various numbers $N$ of the disks. We conducted 20 trials for each value of $N$. Figure \ref{appendix} represents the computation time vs $N$ on a double log-scale, together with a linear fit. The slope of the line is approximately $\alpha=1.91$, which is very close to $\alpha=2$.

\bibliography{references}

\end{document}